\documentclass[11pt]{article}
\usepackage{graphicx, fullpage,empheq}
\usepackage{amsfonts}
\usepackage{amssymb}
\usepackage{mathtools}
\usepackage{amsmath,cases}
\usepackage{indentfirst}
\usepackage{subeqnarray,hyperref,textcomp}
\usepackage{colortbl} 
\usepackage{xcolor} 
\hypersetup{
     colorlinks   = true,
     citecolor    = blue
}

\newcommand{\D}{\textbf{D}}
\newcommand{\x}{\textbf{x}}

\renewcommand{\u}{\textbf{u}}

\newcommand{\bsigma}{\boldsymbol{\sigma}}

\newcommand{\bkappa}{\boldsymbol{\kappa}}

\newcommand{\enz}{\scriptscriptstyle \mathrm{enz}}
\newcommand{\drug}{\scriptscriptstyle \mathrm{dna}}

\newcommand{\inj}{\scriptscriptstyle \mathrm{inj}}
\newcommand{\tot}{\scriptscriptstyle \mathrm{tot}}

\newcommand{\cells}{\textcentoldstyle}
\newcommand{\mcells}{\text{\cells}}
\newcommand{\ECM}{\mathcal{E}}
\newcommand{\Dp}[2]{\frac{\partial #1}{\partial #2}}
\renewcommand{\phi}{\varphi}
\newcommand{\gv}[1]{\ensuremath{\mbox{\boldmath$ #1 $}}} 
\renewcommand{\div}[1]{\gv{\nabla} \cdot #1} 
\newcommand{\nabladiv}[1]{\nabla \cdot #1} 
\newcommand{\paraA}{$T_{Hy}$\,}
\newcommand{\paraB}{$\Delta_T$\,}
\newcommand{\paraC}{$T_{DNA}$\,}
\newcommand{\paraD}{$D_{inj}$\,}
\begin{document}
\title{Numerical Optimization of Plasmid DNA Delivery Combined with Hyaluronidase Injection for Electroporation Protocol}
\author{Daniele Peri$^{(1)}$, Manon Deville$^{(2)}$, Clair Poignard$^{(2)}$, Emanuela Signori$^{(3)}$, Roberto Natalini$^{(1)}$}
\date{\small $^{(1)}$:CNR-IAC -- National Research Council, 
      Istituto per le Applicazioni del Calcolo "Mauro Picone"
      Via dei Taurini 19, 00185 Rome, Italy \\
       $^{(2)}$:Team MONC, INRIA Bordeaux-Sud-Ouest, Institut de Math\'ematiques de Bordeaux, CNRS UMR 5251 \& Universit\'e de Bordeaux, 351 cours de la Lib\'eration, 33405 Talence Cedex, France \\
       $^{(3)}$:CNR-IFT -- National Research Council - Istituto di Farmacologia Traslazionale, Via Fosso del Cavaliere 100, 00133 Rome, Italy \\
       \texttt{d.peri@iac.cnr.it} and \texttt{emanuela.signori@ift.cnr.it}}

\maketitle
\begin{abstract}

The definition of an innovative therapeutic protocol requires the fine tuning of all the involved operations in order to
maximize the efficiency. In some cases, the price of the experiments, or their duration, represents a great obstacle and
the full potential of the protocol risks to be reduced or even hidden by a non-optimal application.

The implementation of a numerical model of the protocol may represent the solution, allowing a systematic exploration
of all the different alternatives, shedding the light on the most promising combination and also identifying the key
elements/parameters.

In this paper, the injection of a plasmid, preceded by a hyaluronidase injection, is simulated through a mathematical model.
Some key elements of the administration protocol are identified by means of a mathematical optimization procedure, maximizing
the efficacy of the therapy. As a side effect of the extensive investigation, robust solutions able to reduce the effects of
human errors in the administration are also obtained.

\end{abstract}
\section{Introduction}

DNA delivery consists in injecting engineered DNA plasmid vectors carrying nucleotide sequences
coding therapeutic molecules, so that transfected cells can work as factory to produce locally
or systematically  specific products to correct pathological defects. It has a deep potential in
revolutionizing therapeutic treatments in the field of infectious and cancer diseases, as
demonstrated in past and recent studies~\cite{wolff,andrereview,Leguebe2017}. However, DNA
transfer is still very limited into the clinical practice: despite the treatment efficacy is
well known, further improvements of delivery conditions are needed. 

An important limitation to DNA transfection is represented by the transport of a plasmid through
the Extra-Cellular Matrix (ECM) and its ability to cross the cell membrane and arriving into the
cell nucleus for its expression \cite{Notarangelo2014}. ECM consists of structural
collagen network embedded in a gel of Glycosaminoglycans (GAGs) and proteoglycans, which prevents
the free diffusion of macromolecules, such as plasmid vectors, and slows down the free diffusion
of cytotoxic drugs, thanks also to the presence of degradation enzymes called
nucleases~\cite{bureau}.

Lot of bioengineering strategies are under investigation for encompassing these barriers. One of
them is the administration of hyaluronidase - an enzyme able to digest the ECM - before the DNA
injection so improving the plasmid distribution within the injected tissue~\cite{buhren,girish}.
This methodology has been employed alone or in combination with electroporation
(EP)~\cite{Derobertis}. The role of hyaluronidase in DNA immunization protocols by electroporation
has been disscussed in \cite{Chiarella2013a, Chiarella2013b, Chiarella2014}. 

EP consists in the application of electric fields to the cell membrane allowing its permeabilization
to favor the DNA internalization~\cite{andrereview,aihara,rols}. Hyaluronidase treatment followed by
EP of plasmid DNA, strongly improves the gene expression~\cite{signori,akerstrom,schertzer}. Although
in~\cite{signori} differences in gene expressions due to time interval between
intramuscular DNA injection and application of electric fields was briefly investigated, lot
remains to be clarified.

Aim of this work is twofold: from one side, we need to identify the correct timing between
hyaluronidase and DNA injection; as a second task, the effects of the waiting time between DNA
injection and EP need to be evaluated and quantified. The use of numerical optimization algorithm,
based on the results of the numerical simulation of the full process (apart from the EP itself),
can help in the production of a large number of alternatives, restricting the range of variation
of the single phases of the therapeutic protocol. The outcome of this study is the maximization of
the effect of the therapy, here represented by the effective area reached by the therapeutic agent
at the time of the application of the EP. 

Paper is organized as follows: a section is devoted to the description of the mathematical model, previously introduced in \cite{Deville2018}. Then, the numerical approach adopted for the determination of the most favorable conditions for
the medical protocol are described, using metamodel interpolation, and the main results obtained by the optimization
procedure are illustrated and discussed. The paper ends with some conclusions and future
perspectives.

\section{Model statement}
\begin{figure}[ht!]
\includegraphics[width=0.7\linewidth]{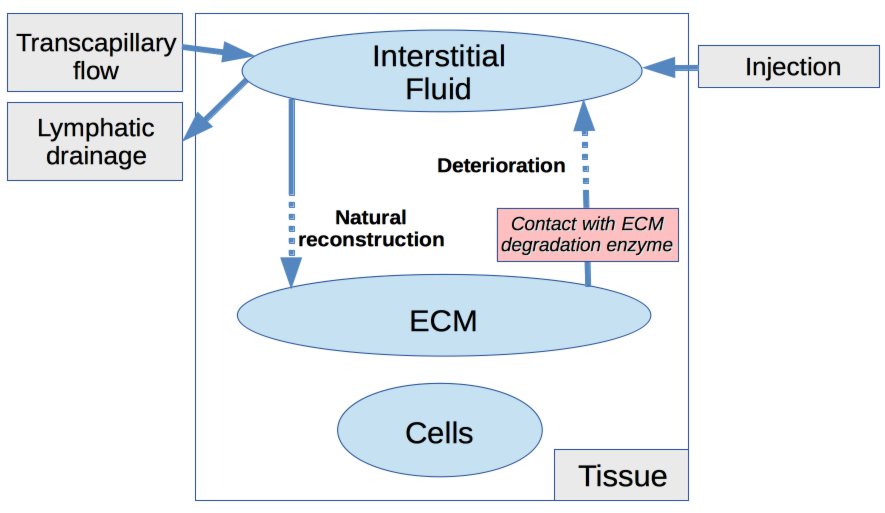} 
\caption{Schematic description of exchange pathways and production terms of the different phases. From~\cite{Deville2018}}
\label{schemaprod}
\end{figure}

In this section, we present the enzyme-based tissue degradation model proposed in \cite{Deville2018}. The model combines the poroelastic theory of mixtures with the transport of enzymes and DNA plasmid densities in the extracellular space. The effect of the matrix degrading enzymes on the tissue composition and its mechanical response are also accounted for. 
The rationale of the model is schematically described in Figure~\ref{schemaprod}.
	
The governing equations are set in the fixed reference domain --the tissue at the initial time-- denoted by $\Omega_0$. For the sake of simplicity, we  assume that our system undergoes very small perturbations (see~\cite{Deville2018} for more details).

The  poroelastic model of Deville~{\it et al.} describes the behavior of the volume fractions of  ECM, cells and fluid --namely the blood in the tissue-- denoted respectively  by $g_{\ECM}$, $ g_{\mcells}$ and $f$, as well as the evolution of the hyaluronidase concentration $h$ and the DNA concentration denoted by $c$. The displacement vector, due to fluid injection is denoted by ${\bf u}$, and $P$ is the inner pressure within the tissue.

The dimensionless model reads as
	\begin{subequations} \label{Bid}
\begin{empheq}[left={\empheqlbrace\,}]{align}
	& g_{\ECM} + g_{\mcells} + f = 1, \label{eqsaturation} \\
	&\div \left((g_{\ECM} + g_{\mcells}) \left(\overline{\lambda} (\nabladiv \textbf{u}) I + \overline{\mu} (\nabla \textbf{u} + \nabla \textbf{u} ^T) \right) \right)= \nabla P, \label{equ} \\
	\begin{split}
	&(g_{\ECM} + g_{\mcells}) \overline{s_0} \Dp{P}{t} - \nabladiv ( \overline{\bkappa} \, \nabla P)= \alpha \, Q_{\inj}^{\tot} + \overline{\gamma}(\overline{P_{v}}-P)\\
	&\hspace{1.5cm}+ \left( \frac{\rho_s^{R,0}}{\rho_f^R} - 1 \right) g_{\ECM} (\overline{K}h + \overline{a_r}(f^{\rm phys} - f) ), 
	 \end{split}
	 \label{eqp}\\
	&\Dp{h}{t} = \nabladiv (f \overline{\D_{\enz}^0} \nabla h+h J_{\enz})  - \frac{\overline{k^d_{\enz}}}{f}h + \frac{\alpha \mathcal{S}_{\enz}}{c_0} , 
	\label{eqh}\\
	&\Dp{g_{\mcells}}{t} + \left( \overline{s_0} \Dp{P}{t}  \right) g_{\mcells}  =  0, \label{eqgcells}\\
	&\Dp{g_{\ECM}}{t} + \left( \overline{K}h + \overline{a_r}(f^{\rm phys} - f) + \overline{s_0} \Dp{P}{t}  \right) g_{\ECM} = 0, \label{eqgECM}
		\\
		&\Dp{c}{t} = \nabladiv (f \overline{\D^0_{\drug}} \nabla c+c J_{\drug})  - \frac{\overline{k^d_{\drug}}}{f}  c+ \frac{\alpha  \mathcal{S}_{\drug}}{c_0},
		\label{eqdrug}
	\end{empheq}
	\end{subequations}
where the fluxes of enzyme and DNA densities are defined by
	\begin{equation}
		J_{\enz} = \frac{1}{f} \overline{\bkappa} \,  \nabla P - \overline{\D_{\enz}^0}  \nabla f 
		\hspace{5mm} \text{ and } \hspace{5mm}
		J_{\drug} = \frac{1}{f} \overline{\bkappa} \, \nabla P - \overline{\D^0_{\drug}}  \nabla f.
		\label{vitconvect}
	\end{equation}
The above partial differential equations (PDEs) system is complemented with initial and boundary conditions.	
Denote by $\Gamma$ the boundary of the domain $\Omega_0$. We generically denote by \textbf{n} the normal to $\Omega$ outwardly directed from the inside to the outside of the domain. We suppose that $\Gamma$ is split into 2 parts denoted respectively by $\Gamma_u$ and $\Gamma_t$ (see Fig.~\ref{Griglia}). 
The following  boundary conditions are imposed 
	 \begin{align}
	& \textbf{S}_s^{E} \, \mathbf{n} = 0
 		\text{ on } \Gamma_t,
\text{ and }
 		\u = 0 
		\text{ on } \Gamma_u
 		\label{BCu}
 		\\
	&	P = 0
			\text{ on } \Gamma_t
	 		\text{ and }
	 		\nabla P \cdot \mathbf{n} = 0
	 		\text{ on } \Gamma_u,
			\label{BCp}
			 \end{align}
		\begin{subequations} \label{BCh}
		\begin{empheq}[left={\empheqlbrace\,}]{align}
	 		& h = 0
			\text{ on } \Gamma_t, \\
	 		& \left(f \D_{\enz}^0 \nabla h + h J_{\enz} \right) \cdot \textbf{n} = 0 \text{ on } \Gamma_{u}.
		\end{empheq}
		\end{subequations}
The same type of boundary conditions are applied to the concentration of DNA plasmid and to its flux.

		\begin{subequations} \label{BCc}
		\begin{empheq}[left={\empheqlbrace\,}]{align}
	 		& c = 0
			\text{ on } \Gamma_t, \\
			&\left(f \D^0_{\drug} \nabla c + c J_{\drug} \right) \cdot \textbf{n} =0
	 		\text{ on } \Gamma_{u}.
		\end{empheq}
		\end{subequations}
		The initial conditions are given in Table~\ref{parnum}.

We can observe from equation \ref{BCc} how the PDEs system involves a large number of parameters. Being the model dimensionless, it is important to recall the link between the dimensionless (with a overline) and the physical (without overline) parameters as given in ~\cite{Deville2018}. We denote by $l_0$ the characteristic length of the tissue.
The dimensionless Piola-Kirchhoff and Cauchy stress tensors are defined as $\overline{\textbf{S}^E_{s}} = \textbf{S}^E_{s}/(\lambda + 2 \mu)$ and $\overline{\bsigma^E_{s}} = \bsigma^E_{s}/(\lambda + 2 \mu)$, respectively, and we define the dimensionless parameters
	\begin{center}
	\begin{tabular}{llll}
	 $\displaystyle \overline{\mu} = \frac{\mu}{\lambda + 2 \mu}$, 
	 & $\displaystyle \overline{\lambda} = \frac{\lambda}{\lambda + 2 \mu}$, 
	 & $\displaystyle \overline{s_0} = s_0 (\lambda + 2 \mu)$, 
	 & $\displaystyle \overline{\bkappa}=\frac{1}{\kappa} \bkappa$, 
	 \\[5mm]
	$\displaystyle \alpha = \frac{l_0^2}{\kappa (\lambda + 2 \mu)}$,
	& $\displaystyle \overline{K}=\alpha c_{0} K$,
	&$\displaystyle \overline{a_r}=\alpha a_r$,
	& $\displaystyle \overline{\gamma} = \frac{l_0^2}{\kappa} \gamma$,	
	\\[5mm]
	$\displaystyle \overline{\D_{\enz}^0} = \frac{1}{\kappa (\lambda + 2 \mu)} \D_{\enz}^0$,
	& $\displaystyle \overline{k^d_{\enz}} = \alpha k^d_{\enz}$,
	& $\displaystyle \overline{P_{v}} = \frac{P_{v}}{\lambda + 2 \mu} $, 
	&	 
	\\[5mm]
	$\displaystyle \overline{\D^0_{\drug}} = \frac{1}{\kappa (\lambda + 2 \mu)} \D^0_{\drug}$,
	& $\displaystyle \overline{k^d_{\drug}} = \alpha k^d_{\drug}$.
	&
	&	 
	\\[5mm]  
	\end{tabular}
	\end{center}
	
We choose the $(\lambda + 2 \mu)$ parameter as a natural pressure scale; by this choice the dimensionless elastic parameters
$\overline{\lambda}, \overline{\mu}$ are of order $1$ \cite{lang}. Some of the physical parameters can be found in the literature,
and others depend on the experimental protocol. The physical parameters that are considered fixed are given in Table~\ref{parnum},
where, if nothing is reported, correspond to the values proposed~\cite{Deville2018}.

The aforementioned systems of equations have been numerically solved by adopting the finite element method: the practical implementation
has been obtained by using the open-source libraries {\tt FreeFEM}~\cite{FreeFem}. Details are provided in~\cite{Deville2018}.

\begin{table}[ht!]
	\tiny
			\caption{Values of the physical parameters fixed for the numerical parametric studies, see also \cite{Deville2018}.}
			\label{parnum}
			\centering
		\begin{tabular}{lllll}
		  \hline\noalign{\smallskip}
		  Parameter & Symbol & Value & Unit & Reference \\
		  \noalign{\smallskip}\hline\noalign{\smallskip}
		  Typical length & $l_0$ & $10^{-2}$ & m &\\
		  Reference concentration & $c_0$ & $10^{9}$ & kg/$\text{m}^3$ &\\
		  Density of fluid phase & $\rho^R_f$ & $10^{3}$ & kg/$\text{m}^3$ &\cite{weiyao}\\
		  Density of solid phase & $\rho^{R,0}_s$ & $1.09 \times 10^{3}$ & kg/$\text{m}^3$ &\cite{ward}\\
		  Specific storage coefficient & ${s_0}$ & $10^{-6}$ & $\text{Pa}^{-1}$ &\\
		  Injected concentration & $c_{\inj}^{\enz}$ & $4 \times 10^{-2}$ & U/$\mu$l & \cite{signori}\\
		  \noalign{\smallskip}\hline\noalign{\smallskip}
		  Permeability & ${ \bkappa}$ & $10^{-11}$ & $\text{m}^{2} \text{Pa}^{-1} \text{s}^{-1}$ & \cite{swartz}\\
		  Lam\'{e} first parameter & ${\lambda}$ & $7.14 \times 10^{5}$ & Pa & \cite{zollner}\\
		  Lam\'{e} second parameter & ${\mu}$ & $1.79 \times 10^{5}$ & Pa & \cite{zollner}\\
		  \noalign{\smallskip}\hline\noalign{\smallskip}
		   Diffusion coefficient of the enzyme & ${\D^0_{\enz}}$ & $10^{-4}$ & $\text{m}^2$/s &\\
		  Diffusion coefficient of the therapeutic agent  & ${\D^0_{\drug}}$ & $10^{-9}$ & $\text{m}^2$/s &\\
		  Starling's coefficient & ${\gamma}$ & $5 \times 10^{-5}$ & $\text{Pa}^{-1} \text{s}^{-1}$ & \cite{soltani}\\
		  Fluid/solute coefficient & $\gamma_c$  & $0.9$ & - & \cite{baxterjain}\\
		  Measure of treatment efficacy & ${K}$ & $10^{-14}$ & $\text{m}^{3} \text{s}^{-1} \text{U}^{-1} $ &\\
		  Recovery coefficient & ${a_r}$ & $5 \times 10^{-4}$ & $\text{s}^{-1}$ &\\
		  Degradation rate of the enzyme & ${k^d_{\enz}}$ & $1 \times 10^{-4}$ & $\text{s}^{-1}$ &\\ 
		  Degradation rate of the therapeutic agent & ${k^d_{\drug}}$ & $2 \times 10^{-4}$ & $\text{s}^{-1}$ &\\
		  Driving pressure & ${P_{v}}$ & $10^{-1}$ & Pa &\\
		  \noalign{\smallskip}\hline\noalign{\smallskip}
		  Initial values & Symbol & Initial value & Unit & \\
		  \noalign{\smallskip}\hline\noalign{\smallskip}
		  Volume fraction of fluid & $\phi_f(0,\x)=\phi^{\rm phys}_f$ & $0.1$ & - &\\
		  Volume fraction of ECM & $\phi_{\ECM}(0,\x)$ & $0.4$ & - &\\
		  Volume fraction of cells & $\phi_{\mcells}(0,\x)$ & $0.5$ & - &\\
		  Network dilatation & $\nabla \cdot \u (0,\x) $ & $0$ & - &\\
		  Enzyme concentration & $h(0,\x)$ & $0$ & $\text{Um}^{-3}$ &\\
		  DNA concentration  & $c(0,\x)$ & $0$ & $\text{Um}^{-3}$ &\\
		  Initial pressure & $p(0,\x)$ & $0$ & Pa &\\
		  \noalign{\smallskip}\hline
		\end{tabular}
		\end{table}

\section{Parametric investigation and optimization}

A mathematical model, describing the different phases of the dosing regimen,
represents a strong and powerful tool for the determination of the correct
execution of the different actions to be taken during the administration protocol.
In particular, starting form the analysis produced in \cite{Manon}, we have observed
how some parameters are not optimally selected, although they appears to be able to change
deeply the final effect of the whole procedure. Some experimental trials have been 
produced in order to drive the selection of the best values, but the number of attempts
is clearly limited by the costs of the experimental activity, and the final result can be
reasonably further improved. Under this perspective, the use of a mathematical model would
be of great aid.

\begin{figure}[h!]
\begin{center}
\includegraphics[width=0.95\textwidth]{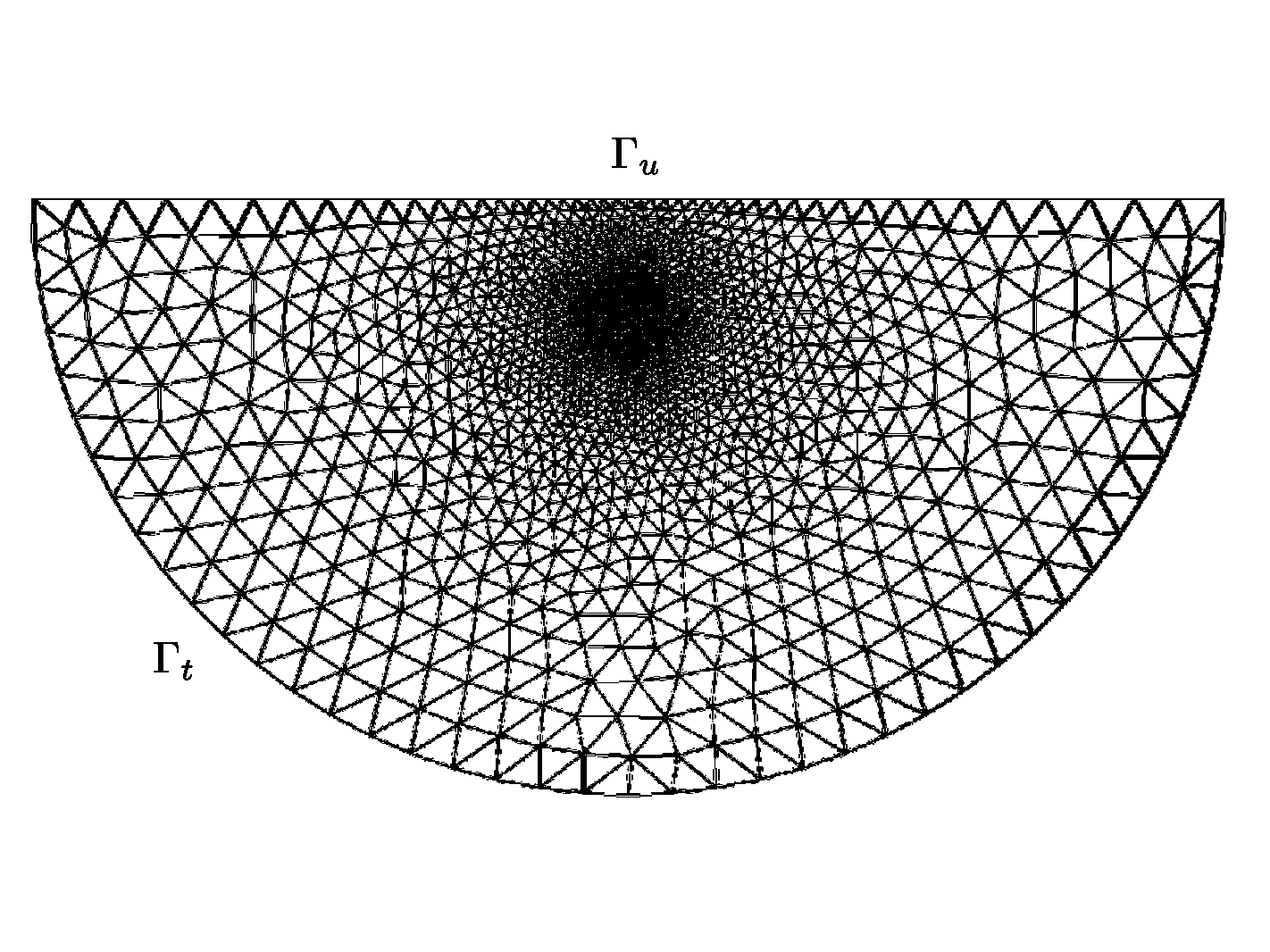}
\caption{Computational grid for the current problem: the density of the cells is increased in the
         area around the injection point.
         }\label{Griglia}
\end{center}
\end{figure}

The numerical simulation of the effects of the complete dosing regimen is obtained by discretizing
a portion of the tissue where the injections will be performed and observing the diffusion of the
plasmid and the DNA in this volume, resolving the previously described systems of equations. In
figure \ref{Griglia}, an example of the computational mesh is reported. The problem is solved
numerically under the assumption of spherical symmetry: for this reason, the solution of a
two-dimensional representation of the tissue is sufficient. In the picture, the semi-circular area
represents a planar section of an half-sphere. The upper border (planar) represents the skin
surface, while the semi-circular area represents a portion of the tissue under the skin surface.
The diffusion of the therapeutic agents is observed into this volume.

\subsection{Selection of relevant parameters}

In order to find the best value of the different parameters, they can be systematically changed: a
number of different configurations are numerically evaluated with the aim of detecting the best
strategy for the dosing of the therapy. To do that, we have to define a criterion for the quantification
of the preference of a configuration with respect to another: reasonably, we can
assume the maximum value of the volume occupied by the DNA during the time as a measure of the
effectiveness of the combination of parameters. Considering that a minimal quantity of the plasmid
is required in order to be effective, we can assume the area where the concentration of the
plasmid is higher than a minimum value as the effective area. Since a deterioration of the plasmid
is observed in time, the effective area is typically growing during the injection phase and in the 
successive moments, but after a certain time (depending on the administration strategy) it is
decreasing. With the mathematical model we are able to take trace of this evolution, determining
its maximum value and the time at which it occurs, including also the evolution of the effective
area after the maximum value has reached. The plasmid is here considered as effective when it
represents at a concentration of 5\% or more: this check is performed on every cell adopted in the
discretization of the investigated volume, and the area of every active cell is contributing to
the full effective area.

\begin{figure}[h!]
\begin{center}
\includegraphics[width=0.95\textwidth]{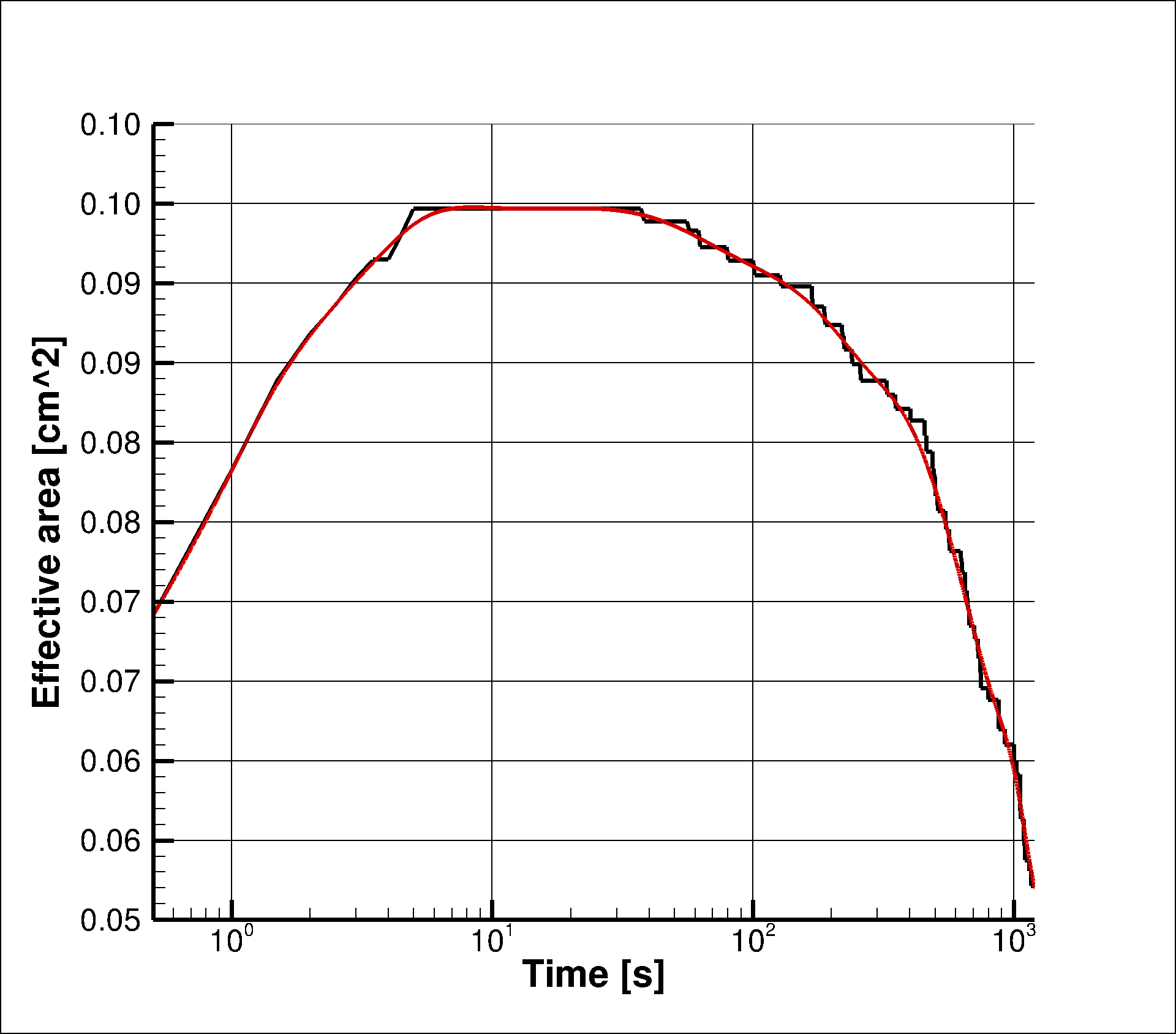}
\caption{Evolution of the area reached by the therapeutic agent during the time starting from
         the end of the plasmid injection.
         }\label{Evo}
\end{center}
\end{figure}

Among the different elements of the dosing regimen, we put our attention on the sequence of the
injections. Here we can observe four different phases:
\begin{itemize}
\item Hyaluronidase injection.
\item Waiting time for the hyaluronidase diffusion into the tissue.
\item Plasmid injection.
\item Waiting time for the plasmid diffusion into the tissue.
\end{itemize}

For each injection, there are some undetermined quantities:
\begin{itemize}
\item The duration of the injection phase.
\item The amount of the injected substance.
\item Deepness of the injection.
\end{itemize}

The injected quantities of hyaluronidase and plasmid have been fixed at $25\mu\,l$ and $30\mu\,l$
respectively, and the deepness of the injections have been assumed to be the same for both. After
these choices, we have still four free parameters:
\begin{enumerate}
\item Duration of the hyaluronidase injection \paraA .
\item Waiting time for the hyaluronidase diffusion into the tissue \paraB .
\item Duration of the plasmid injection \paraC .
\item Deepness of injections \paraD .
\end{enumerate}
These four parameters have been adopted, in the following, as the design parameters whose best
value is searched.

\subsection{Features of the mathematical model}

The determination of the optimal values of the four parameters requires, in general, the application
of an optimization algorithm: once a mathematical programming problem is formulated, a large number
of trial vectors of the design parameters need to be automatically generated and evaluated, as soon
as the convergence to the optimal values of the parameters is obtained. Unfortunately, the numerical
noise connected with the numerical solution of the problem and the large computational time required
to finalize a single simulation represent two great obstacles in the application of this approach.
In fact, a noisy behavior of the function to be minimized/maximized is typically creating a number
of false minima/maxima, and the optimization algorithm is sometime trapped into those regions.
Regarding the CPU time for the solution of a single configuration, depending on the values of the
parameters, it could be greater than five hours, and around ten thousand of
simulation are needed for reaching the convergence to the optimal solution in our case. 

\begin{figure}[h!]
\begin{center}
\includegraphics[width=0.95\textwidth]{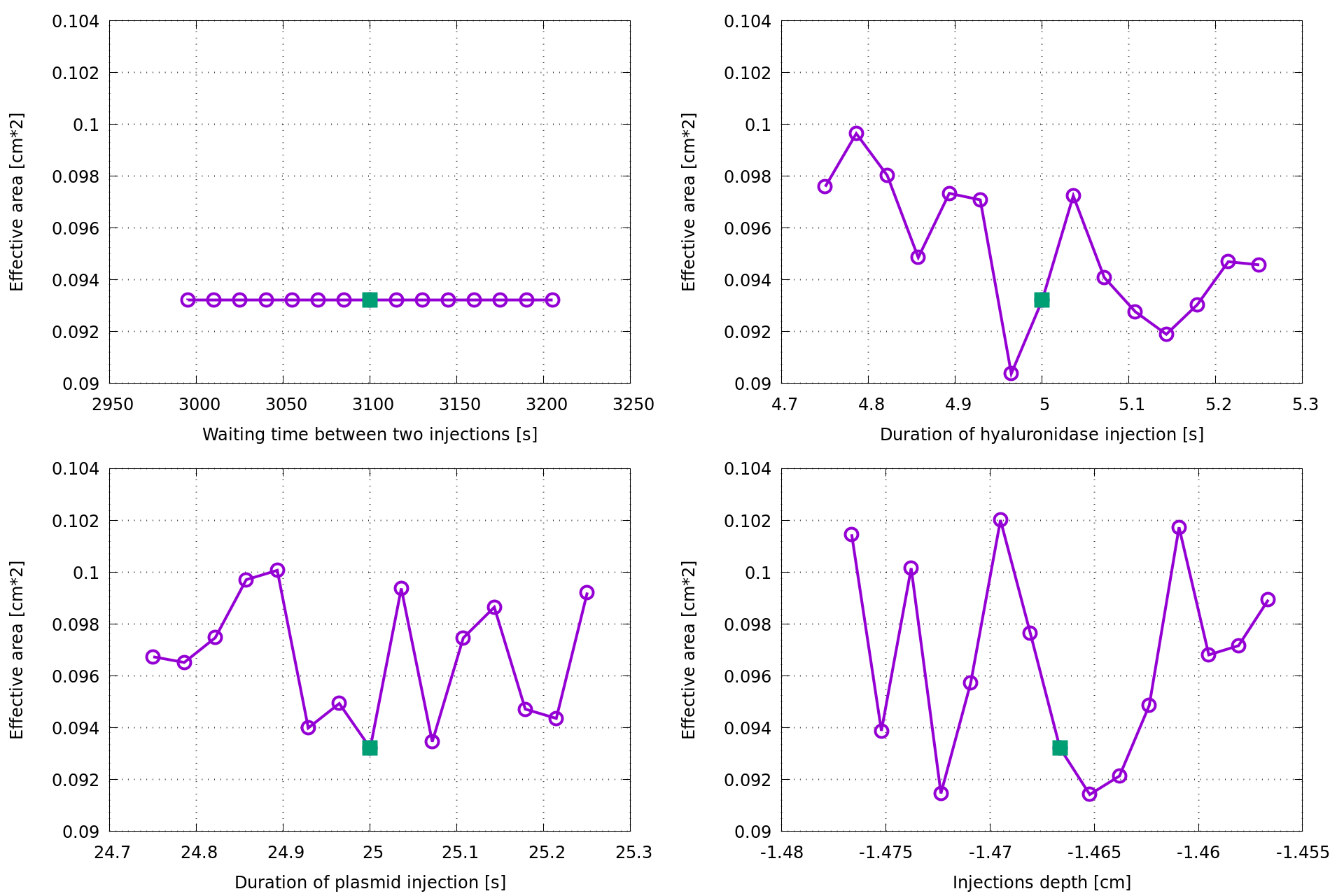}
\caption{Sensitivity analysis of the area reached by the therapeutic agent as a function of
         the injection deepness only. On the left, comparison between the point values and the
         interpolated ones, on the right the interpolated values only (scales are different).
         }\label{Sensi}
\end{center}
\end{figure}

An example of the effects of the numerical noise is reported in figure \ref{Sensi}. Here a very
small variation of a single parameter is enforced, while all the other parameters are kept fixed.
The effect of the variation of a single parameter on the total effective area is reported in the
corresponding sub-figure. The central value is representing one of the best configuration identified
during the following exploration. We can observe how, in the investigated region, the time between
the two injections is not changing at all the value of the effective area, while for the other
parameters a nearly random effect is observed: it is evident that a sort of uncertainty is connected
with the estimate of the effective area, and the simple punctual value provided by the simulation
cannot be representative of the real effects of the selected parameters. For this reason, we
should try to define a different value of the effective area, able to put into consideration the
strong local sensitivity to the parameters of the output of the simulations. We decided here to
apply a worst case approach, and the average value of a group of local samples, reduced by the
associated variance, is defined as our objective function. Statistically, the use of this quantity
guarantees that the effective area in the neighborhood of the selected configuration of the dosing
parameters is greater than the indicated value with a probability of 84.15\%. For this reason, from
now on we will refer to the effective area as its average value (computed on a sampling set of 9
configurations) minus the variance.

\begin{figure}[h!]
	\begin{center}
		\includegraphics[width=0.95\textwidth]{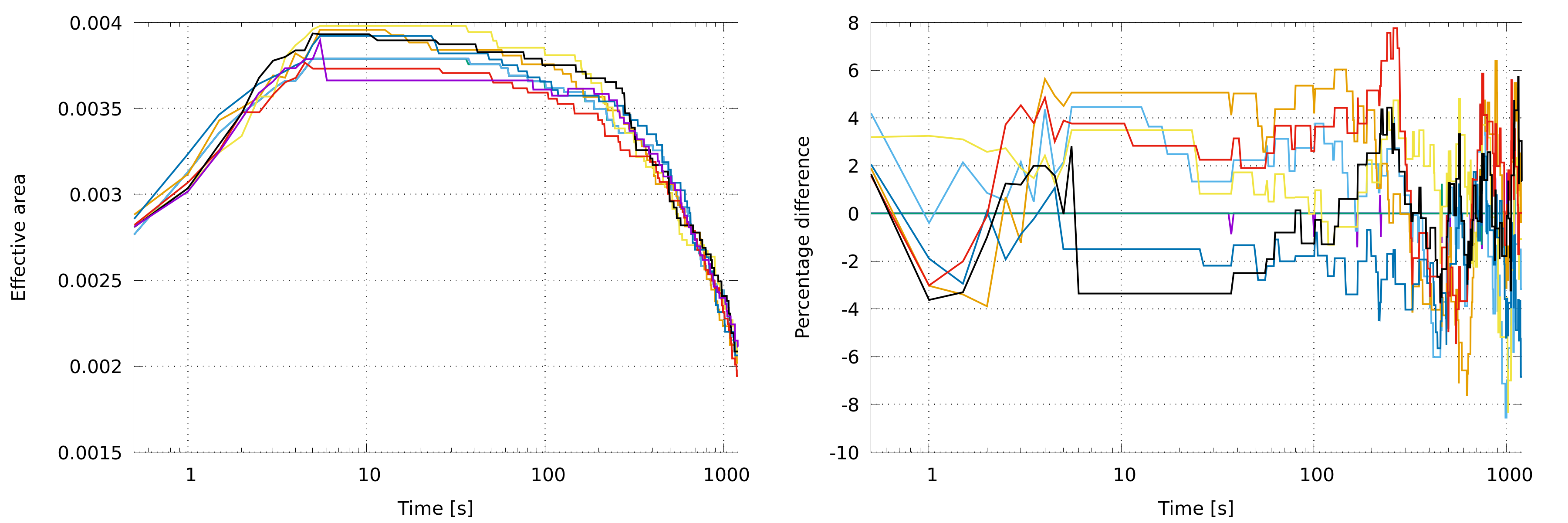}
		\caption{Absolute (left) and percentage (right) differences observed on the nine configurations
                         adopted in the sensitivity analysis. Percentage differences are
		         computed with respect to the central point of the distribution.
		}\label{Unc}
	\end{center}
\end{figure}

In figure \ref{Unc} we have reported the full evolution curve of the effective area for the nine
configurations adopted during the sensitivity analysis of a single simulated point. On left, the
differences in absolute terms are reported, on the right the percentage differences are shown.
Percentage differences are computed with respect to the value of the central point of the
distribution. We can observe how a difference of about eight percentage points is recorded
among the different curves: this represents a sign of great sensitivity of the simulation model
to the parameter variation. The fluctuations are slightly amplified by the lower value of the effective
area at the end of the curve: anyway, this behavior represents a strong element for
the consideration of an averaged value instead of a punctual value of the effective area.

Another aspect that we should take into consideration, and that advises against the use of an
optimization algorithm, is that the practical implementation of the prescription includes human
errors, so that the precision of the effectively realized parameters is practically low. As a
consequence, we are much more interested in some general indications about the order of magnitude
of the various parameters rather than an extremely precise list of values for our parameters.
Human errors can be reduced by a partial or even complete automation of the different operation,
however some steps should be necessarily performed manually, in particular when a living creature
is involved in the application.

\subsection{Selection of the optimization strategy}

On the base of the previous considerations,
the strategy adopted in this paper is based on the general framework of the "Multipoint Approximation Method"~\cite{Toropov1995},
or, more in general, of the "Metamodel-Based Simulation Optimization" (see \cite{Barthelemy1993,Barton2006}).
This approach can be synthetically described as follows. The first step consists in the generation of a suitable number of
{\em training points}, preferably regularly spaced, spanning the design variable space. The number of the training points is
unknown {\em a priori}, depending typically on the rate of variation of the objective function and on the number of design
variables. It can be fixed also considering the allowed computational effort for the optimization activities.
The objective function is now computed at the training points, obtaining the so called {\em training set}.
The training set is used in order to derive an interpolation/approximation of the objective function over the full design
space, generally called {\em metamodel} (a model of the model). The metamodel is substantially an algebraic model, able
to mimic the numerical response of the computationally expensive mathematical model. The training phase of the metamodel depends
on the characteristics of the metamodel itself: during the training phase, the parameters of the metamodel are optimized
in order to minimize the prevision error. Some metamodels are trained easily, i.e. by solving a linear system whose dimension
is equal to the number of training points, other metamodels require the solution of an optimization algorithm (like neural
networks). A small part of the training set can be put momentarily aside, forming the {\em verification set}, and then it can
be used at the end of the training phase in order to verify the accuracy of the metamodel on positions not previously used during
the training. Several techniques can be now adopted in order to select new training points with the aim of increasing the accuracy
of the prediction, if required: examples are reported in~\cite{peri2009self,Shu2016}. Once the quality of the metamodel is
satisfactory, it can be applied to the optimization algorithm, in order to identify the optimal parameters of the mathematical model.
Since the evaluation of the metamodel is computationally inexpensive if compared with the mathematical model, the overall computational
cost of the optimization procedure is equal to the time of the training phase.

\subsection{Details of the optimization strategy}

In this application, we selected the {\em Orthogonal Arrays}~\cite{Sloane}(OA) for the generation of the training set. OA
represents a reduced set with respect to the full factorial design obtained by a regular sampling of each coordinate direction:
some configurations are deleted from the full factorial design, and the remaining ones are respecting an orthogonality
criteria. In this case, 16 levels for the regular subdivision of each direction of the design space have been selected:
a complete full factorial design is composed by 65536 points, here reduced to 512 sampling points after the application of
the OA criterion. Since we need to perform a sensitivity analysis for each sample point, the total number of configurations
to be analyzed is 4608.

As a surrogate model, a multi-dimensional spline approximator \cite{peri2017spline} has been adopted, tuned using the
results provided by the previously produced training set. The interval of variation for the parameters has been fixed
as follows: \paraB is varied between 300 and 10800 seconds; \paraA and \paraC are varied between 5 and 30 seconds,
and \paraD is varied  between 1 and 2 centimeters. These values have been selected observing previous numerical
experiences from \cite{Manon}. 

\begin{figure}[h!]
\begin{center}
\includegraphics[width=0.95\textwidth]{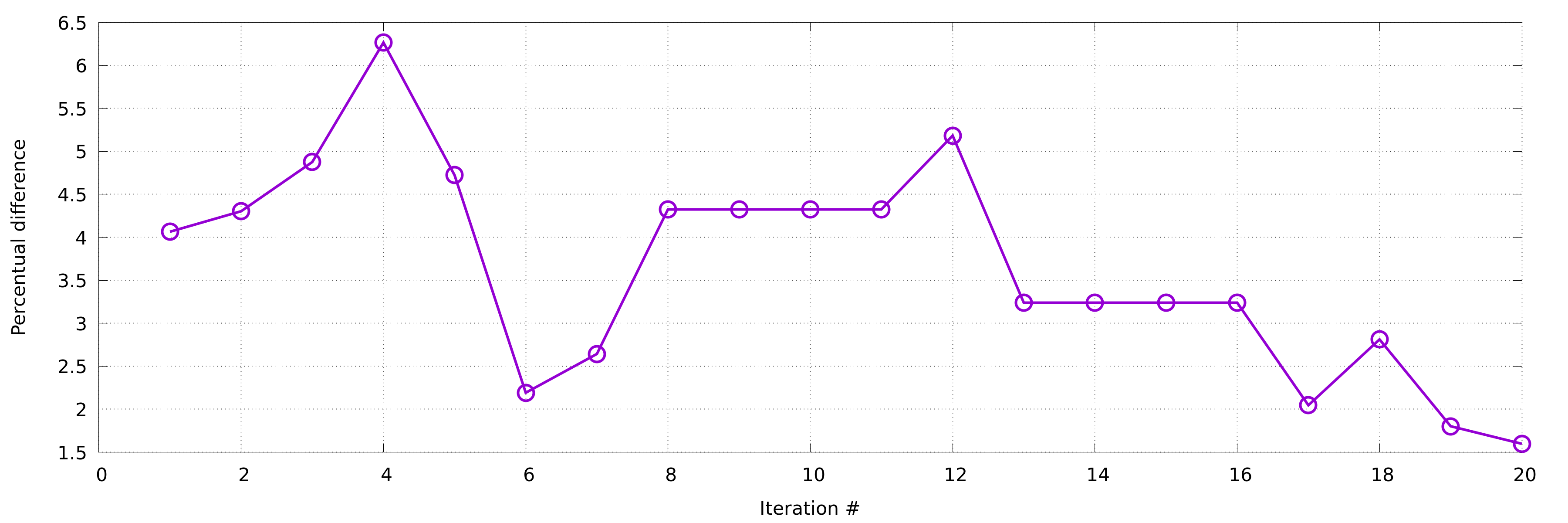}
\caption{Precision index of the meta-model during the refinement phase.
         }\label{Perc}
\end{center}
\end{figure}

In order to increase the credibility of the meta-model, some further training points have been added sequentially in
those areas where the objective function appears to be favorable. The full number of training points, at the end of the
refinement phase, has become 686 (6174 configurations). A solution of mathematical model is produced for the new training
points, and the difference between the value of the objective function estimated by the meta-model and the real value
provided by the mathematical model at the new point is assumed as the precision index of the meta-model. The history
of the refinement phase is reported in figure \ref{Perc}. Range of variations of the parameters are also adjusted.

The determination of the best configuration is obtained by regularly sampling the design space and
then recursively refining the investigation as soon as the dimension of the investigated area is
lower than a prescribed limit. Since this operation is completed by using the metamodel, we can adopt
very strict parameters: we have here 51 subdivision along each coordinate direction and the final
spatial precision of the search is fixed at $10^{-8}$.

\subsection{Results}

So far, the maximum value of the effective area have been considered. We can observe in picture \ref{Evo}
that the best value is typically obtained after ten seconds from the end of the plasmid injection, 
and this value is nearly constant for about 30 seconds: this is in line with the practical necessities of
the preparation of the following EP (normally form 10 to 20 seconds). Since after that time the effective
area is reducing, if the waiting time for the EP is longer than 30-40 seconds we have a loss of efficiency
of the full prescription. In order to estimate the degradation of the effective area in time, 5 different
pictures have been produced at 1, 2, 4, 8 and 16 minutes after the end of the plasmid injection. In this
case, the value of the objective function is reported, so that following pictures are showing the average
of the effective area reduced by their variance. Since we need to take into consideration the inaccurate
realization of the parameters, in the pictures a dot is reported only if the objective function is at least
greater than 99\% of the maximum observed value of the objective function: this way, we have a representation
of the areas of the parameter space where we can substantially guarantee the maximum efficiency of the prescription,
reducing the negative effects of a small error of the implementation. Results are reported for different values
of the depth, that is fixed in each sub-picture at the indicated value, and for two different waiting time before EP,
that is, 1 and 16 minutes.

\begin{figure}[h!]
\begin{center}
\includegraphics[width=0.95\textwidth]{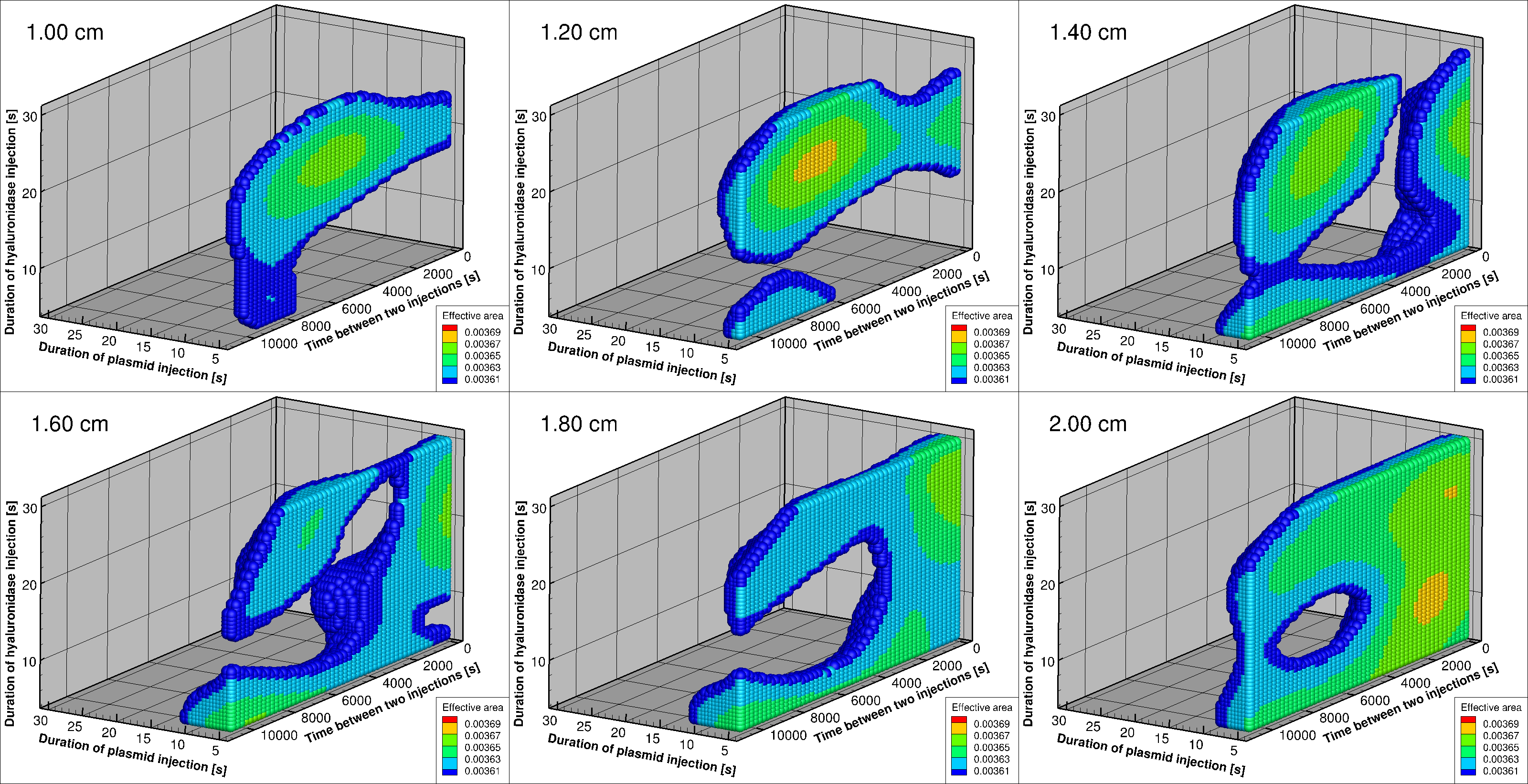}
\caption{Range of parameters for which a loss of 1\% with respect to the maximum realizable effective area is
         obtained. Time after the plasmid injection completion: 1 minute.
         }\label{1min}
\end{center}
\end{figure}

If we are able to complete the preparation of the EP phase in 1 minute, the duration
of the plasmid injection should be very small. From figure \ref{1min} we can observe how the
preferable value of \paraD is not univocal, since similar values are obtained for 1.2
centimeters and 2 centimeters, but the values of \paraB are different,
being shorter in the case of the deeper injection. This may represents an advantage if a series
of experiments are performed.

\begin{figure}[h!]
\begin{center}
\includegraphics[width=0.95\textwidth]{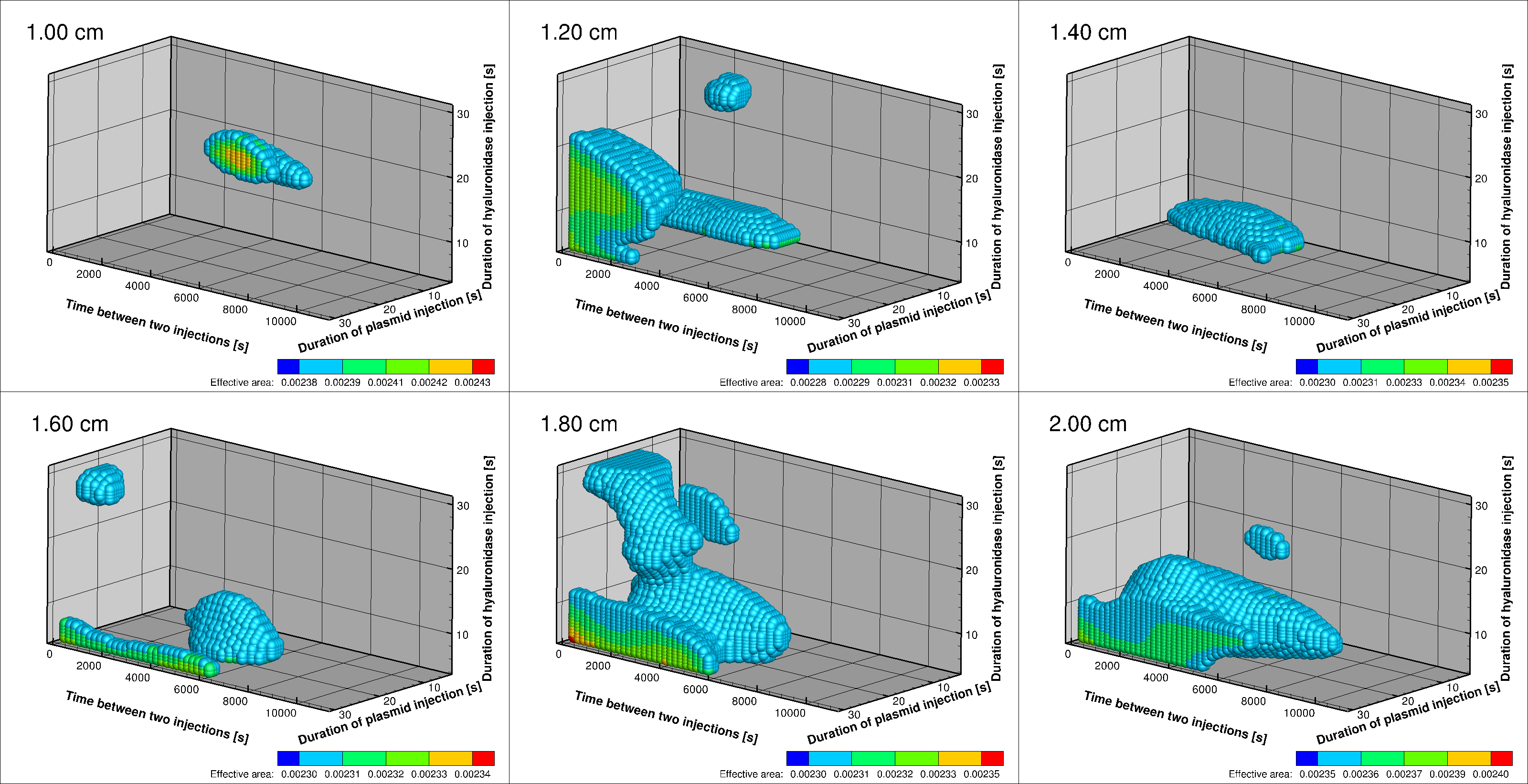}
\caption{Range of parameters for which a loss of 1\% with respect to the maximum realizable effective area is
         obtained. Time after the plasmid injection completion: 16 minutes.
         }\label{16min}
\end{center}
\end{figure}

On the contrary, if we observe the results after 16 minutes, reported in figure \ref{16min}, the value
of the objective function for the case of \paraD equal to 2 centimeters is larger than in
the other cases. \paraC can be longer, while \paraB is variable.

\begin{figure}[h!]
\begin{center}
\includegraphics[width=0.95\textwidth]{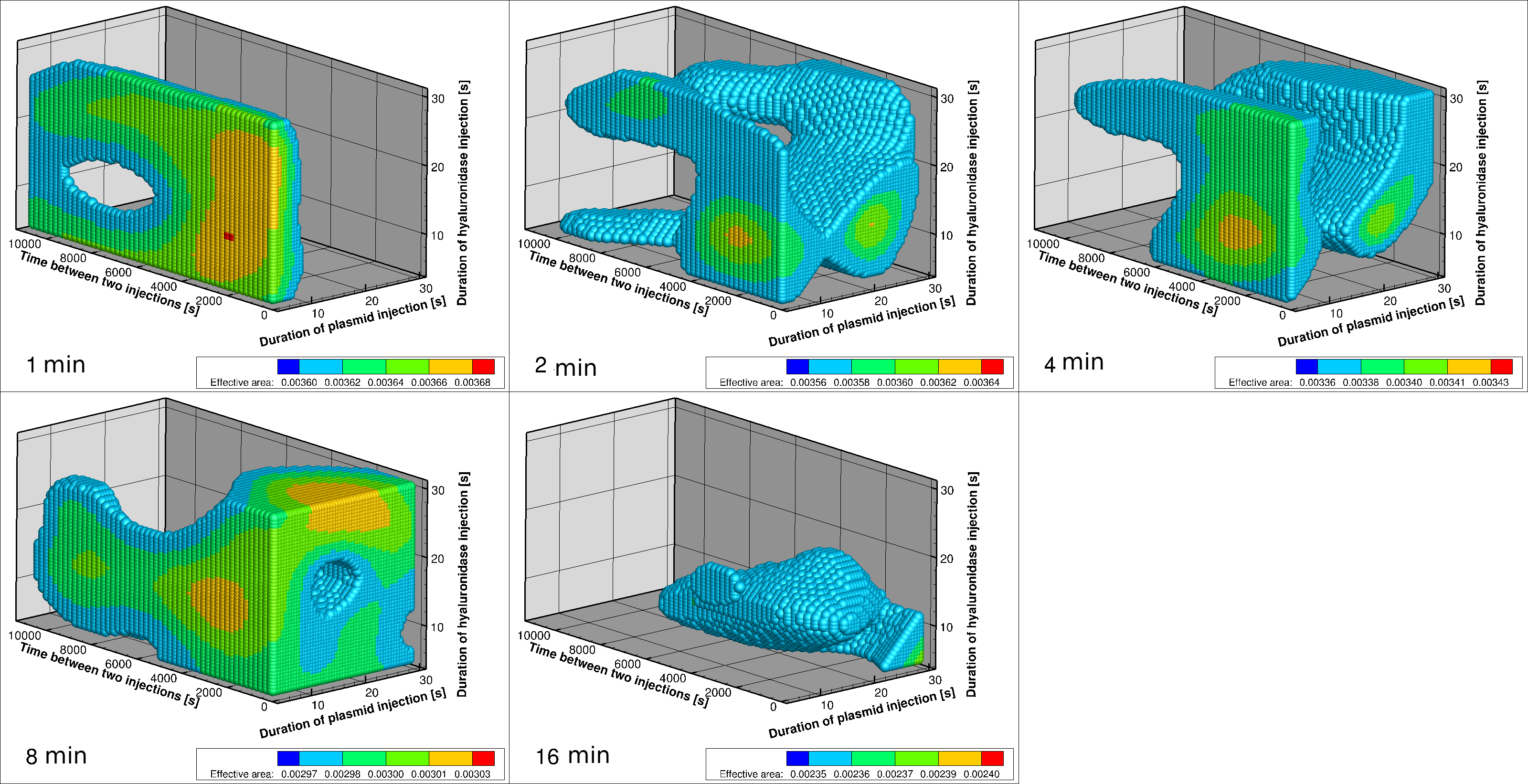}
\caption{Range of parameters for which a loss of 1\% with respect to the maximum realizable effective area is
         obtained. Depth of the injections: 2 centimeters.
         }\label{depth40}
\end{center}
\end{figure}

If we now fix \paraD at 2 centimeters, we can observe the effect of \paraB, reported in figure \ref{depth40}.
The scale of the objective function is changing from sub-picture to sub-picture.

\begin{figure}[h!]
\begin{center}
\includegraphics[width=0.95\textwidth]{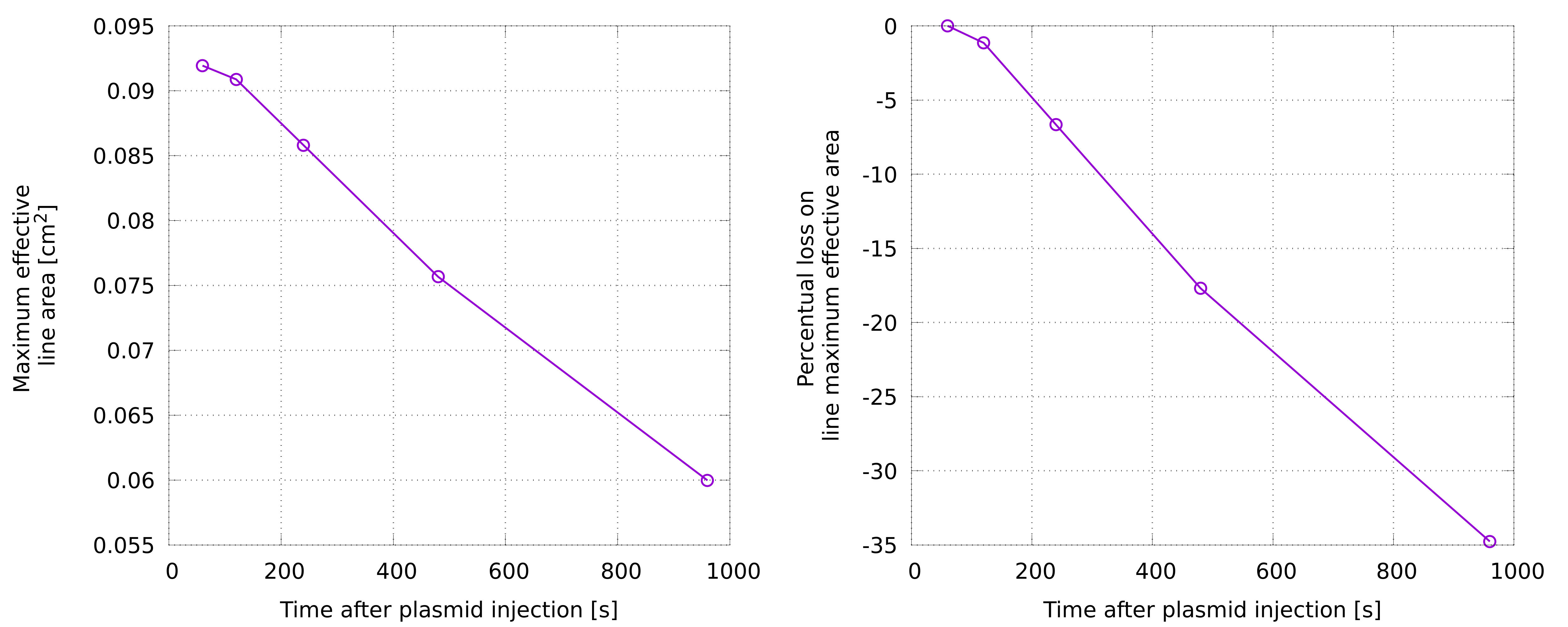}
\caption{Loss in the effective area with an increasing \paraB. On top, absolute values. On bottom, percentage values.
         }\label{m40}
\end{center}
\end{figure}

After 8 minutes we have the larger tolerance in the best parameters: this area is largely reduced
if the EP is performed after 16 minutes. At the same time, the objective function is reduced if
the EP waiting time is increased. The numerical estimate of this loss is reported in figure \ref{m40}.

\begin{figure}[h!]
	\begin{center}
		\includegraphics[width=0.95\textwidth]{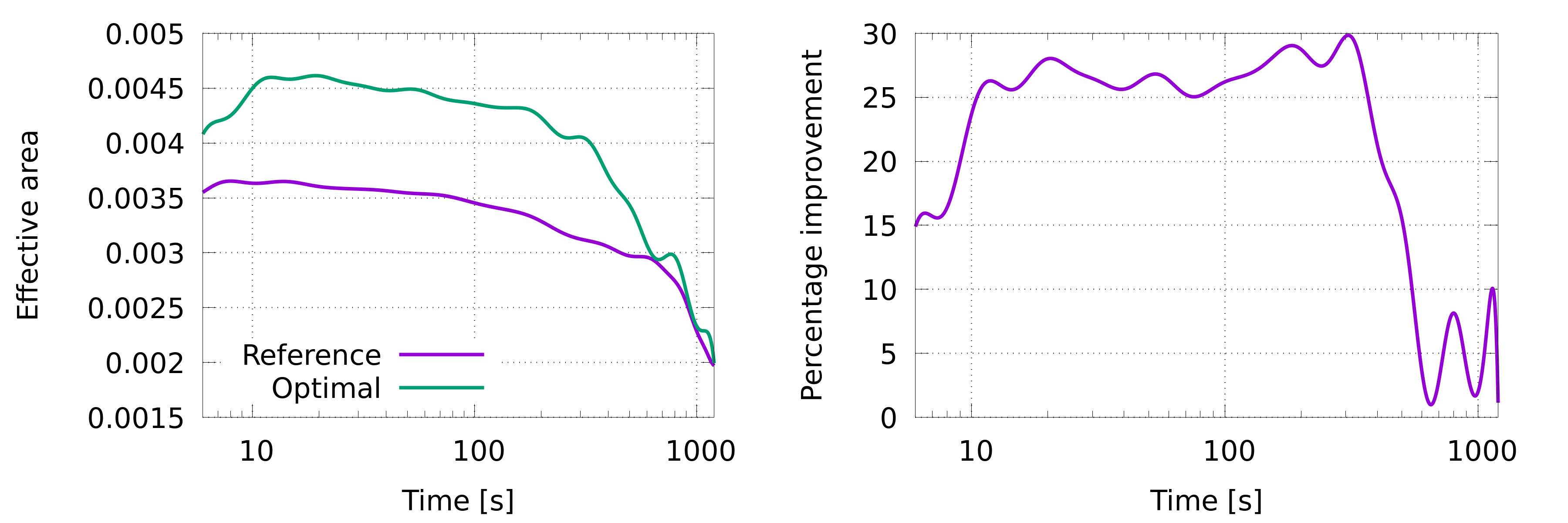}
		\caption{Comparison between a reference configuration and the overall optimal
                         configuration identified by the numerical analysis.
		}\label{rif-opt}
	\end{center}
\end{figure}

In order to quantify the improvements potentially obtained by the optimization procedure, a
reference configuration, commonly adopted for this kind of experiments, has been compared with
the best configuration identified by the optimization procedure. Results are reported in figure
\ref{rif-opt}. An increase of about 30\% is obtained if the EP is performed after
no more then 8 minutes from the DNA injection: in fact, it is obviously convenient to perform the
EP at the moment of the maximum expansion of the DNA into the tissues. If
the waiting time is greater than 10 minutes, the two different strategies does not show
significant differences, probably because the dynamics connected with the degradation of the DNA
are substantially independent from the protocol details. The observation of these results suggest
not to delay the EP procedure after 30 seconds: this request may require the application of a fully
automated operations, in order to reduce the time spent in all the different preliminary sub-activities
required by EP, such as correct immobilization of the subject undergoing to EP, application of
conductive gel in the area to treat, correct placement of the electrodes, etc.). This time reduction
is of paramount importance since a longer waiting time is almost nullifying all the advantages
obtained by the optimized protocol.

\subsection{Role of \paraC}

In the common practice, a quite long waiting time between the two injections, hyaluronidase and
DNA, is adopted. Typically, \paraC is about one or two hours later respect to the hyaluronidase
administration. This assumption has been motivated by the necessity to be sure that the hyaluronidase
has enough time to develop its effect, increasing the porosity of the ECM and then facilitating
the access of the DNA-plasmid at the border of the cells.

\begin{figure}[h!]
     \begin{center}
     \includegraphics[width=0.45\textwidth]{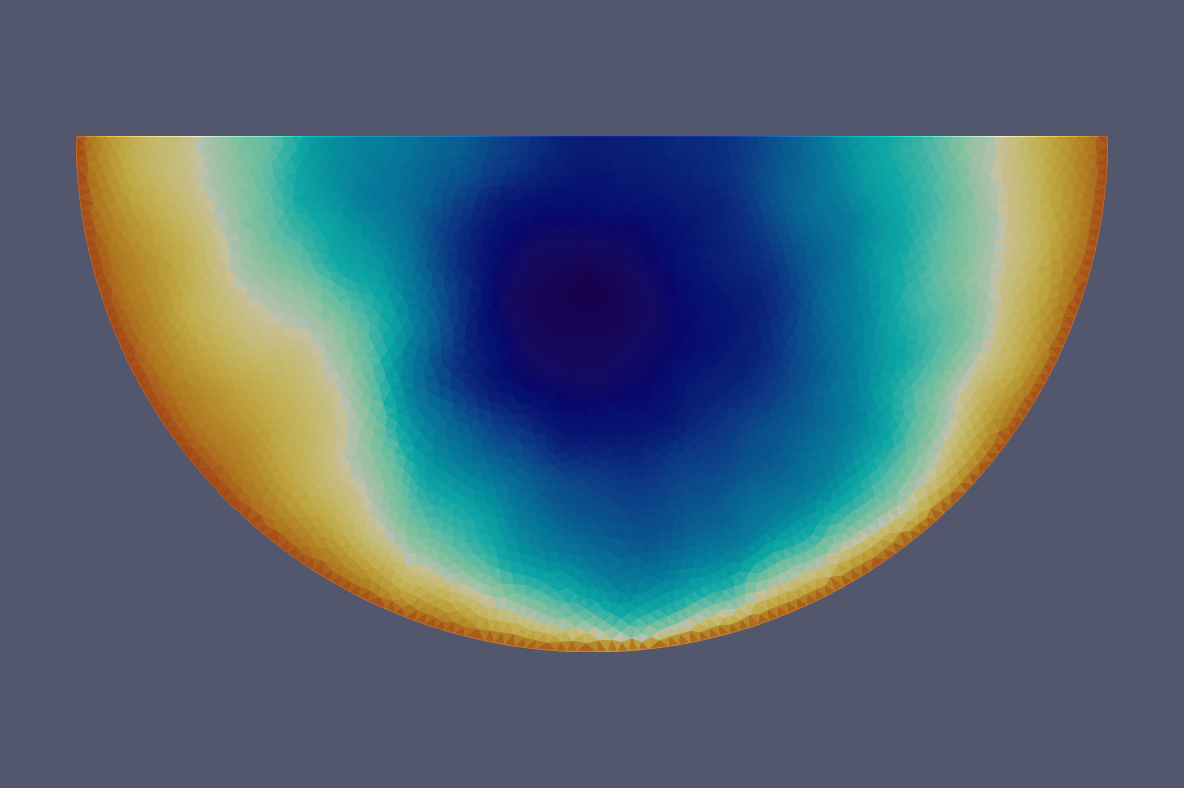}
     \includegraphics[width=0.45\textwidth]{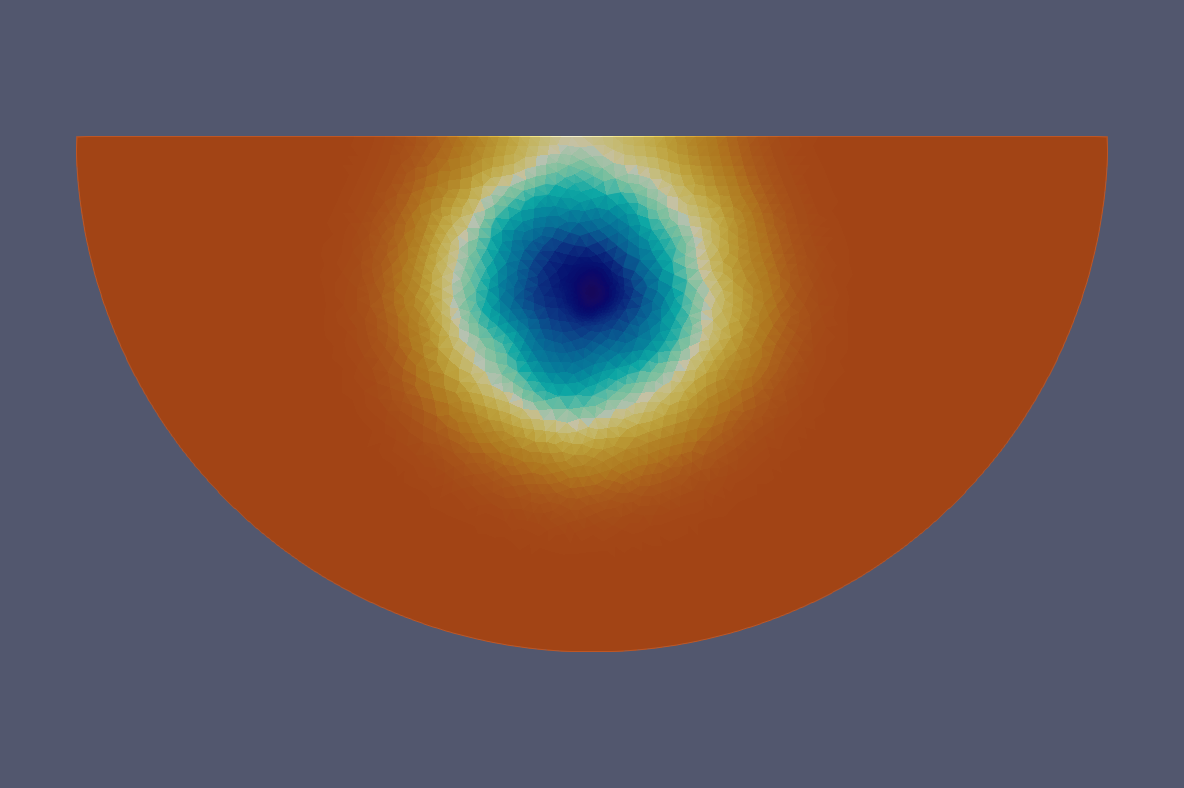} \\
     \includegraphics[width=0.45\textwidth]{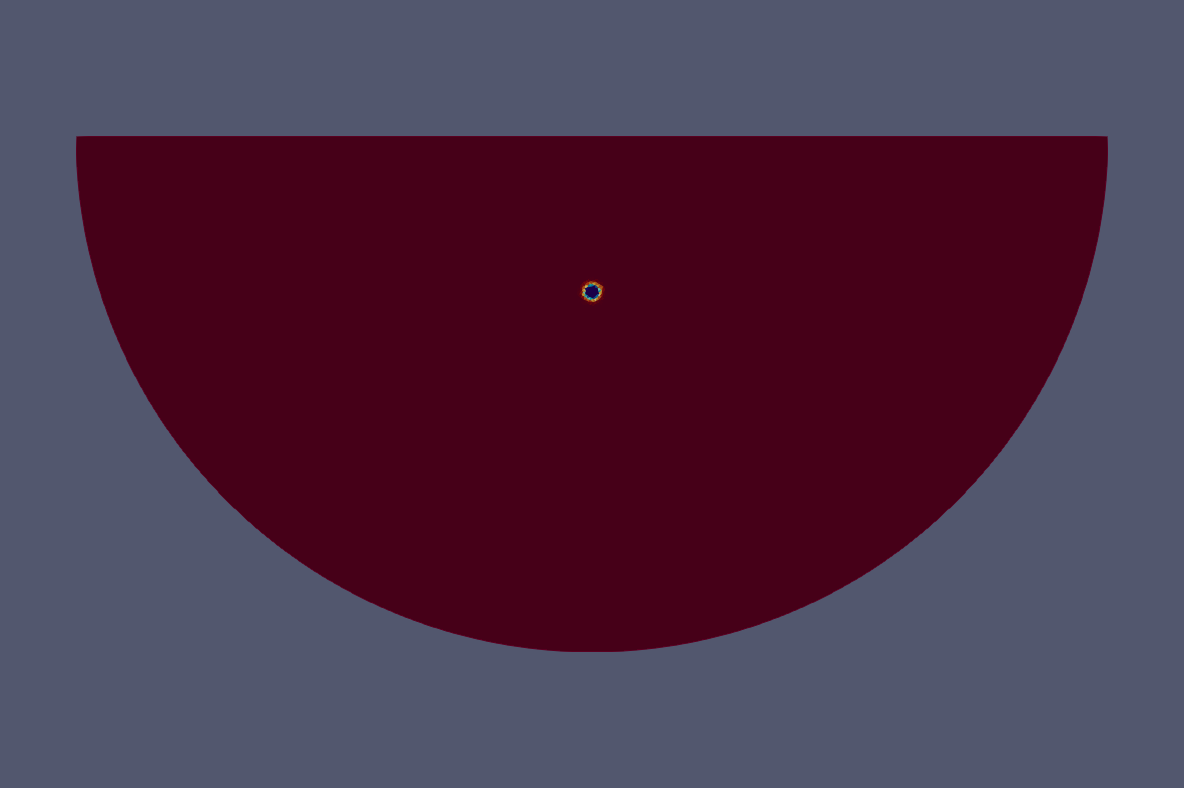}
     \caption{On top: Porosity variation in the computational domain at the corresponding \paraB.
              On left, \paraB is 3800 seconds, on right is about 290 seconds.
              On bottom, the maximum area where the concentration of the DNA-plasmid is larger than 5\%.
             }\label{hyal-time}
     \end{center}
\end{figure}

At the end of the solution of the optimization process, two categories of best solutions have
been identified: the overall best configuration ever computed in terms of effective area and
the best configuration in terms of objective function. The second solution is the preferable
one, since it takes into account the variability of the effective area under small variations
of the control parameters. The set of realized parameters differs largely between the two
solutions: in particular, while in the first case \paraB is 3500 seconds, in the
second case \paraB is nearly 300 seconds. This second option is absolutely preferable, since
it reduces a lot the overall execution time of the experiment, but a deeper analysis is needed
in order to understand why different timings led to similar results. In figure \ref{hyal-time},
the elements for a reasonable explanation are reported. In the picture, we have plotted the porosity
of the tissue at the time of the DNA-plasmid injection for the two different values of \paraB. In
the same picture, the maximum expansion of the DNA-plasmid in the tissue is also reported, plotting
the area in which the concentration of DNA-plasmid is more than 5\%. We can observe how the effect
of the hyaluronidase is clearly increasing with \paraB, since almost the whole computational
volume has been interested by the action of the hyaluronidase when \paraB is 3500 seconds.

For the shorter value of \paraB, the same effect is obtained in the very central part of the computational
volume, but the effect vanishes quickly. On the bottom part of figure \ref{hyal-time}, the area in
which we have a significant concentration of DNA-plasmid is reported: we can observe how this area
is really small, so that the effect of the hyaluronidase in the case of the small \paraB is
absolutely sufficient.

This result is connected with the large dimensions of the DNA-plasmid, and consequently its great
difficulties in traveling into the ECM. Observing this result, we can argue that multiple injections
of DNA-plasmid in a small area could probably increase (linearly with the number of injection sites)
the overall effect of the protocol, since further modifications of the method of administration of the
hyaluronidase appear not to be effective.

In table \ref{parametri}, the exact values of the design parameters are reported for three different
options for the protocol: the reference values ($\alpha$), based on some indications about common
practice adopted in pre-clinical protocols, the configuration providing the best overall value of
the effective area ($\beta$) and
the configuration maximizing the objective function ($\gamma$). Since the objective function is
taking into account the stability of the solution in the neighborhood of the computational point,
this last configuration is preferable. We can observe that, in the case $\alpha$ we have the maximum
values for \paraB and \paraD, while \paraA and \paraC are the smaller ones. Both $\beta$ and $\gamma$
suggest a smaller value of \paraB, significantly shorter in the case $\gamma$. This is probably the more
interesting result, since this allows for a huge compression of the overall duration of the experiment
and gives also some indications about the behavior of the DNA-plasmid. \paraA and \paraC are larger for
both $\beta$ and $\gamma$ with respect to $\alpha$: due to the small injected quantities, probably the
implementing rules suggested in $\beta$ cannot be practically achieved. The smaller value of \paraD
in $\beta$ and $\gamma$, on the contrary, is compatible with the experimental setup.

\begin{table}[ht!]
\caption{Values of the control parameters for three different configuration of the protocol:
         reference values (current common practice), best effective area (punctual value),
         best objective function value (locally averaged value).}
\label{parametri}
\centering
\begin{tabular}{llll}
Parameter   & Reference & Best EA & Best OF \\
\paraA [cm] &    2.00   &    1.47 &    1.52 \\
\paraB [s]  &   10.00   &   28.33 &   21.47 \\
\paraC [s]  & 5400.00   & 3800.00 &  289.60 \\
\paraD [s]  &   10.00   &    9.98 &   14.54 \\
\end{tabular}
\end{table}

\section{Conclusions}

Improvements of gene electrotranfer protocol are becoming of paramount importance to translate
this treatment into human patients. When DNA plasmid vector is injected into tissues, its
expression is limited, due to the presence of ECM and cell membrane barriers. The employment of
hyaluronidase, which allows a partial digestion of the ECM, and EP - a physical
methodology favoring cell membrane permeabilization - represents a valid platform for DNA delivery
expression. In this study, a mathematical model simulating the core part of the delivery protocol
has been applied in order to enhance the DNA expression, identifying the best injection and waiting
time for the DNA administration. Once numerically compared with the standard operative protocol,
the optimal strategy returns an improvement of about 30\% on the DNA delivery expression. This
encouraging result is subject to the capacity of maintain the waiting time between the DNA
injection and the application of EP inside a maximum of 200 seconds.

A dedicated experimental {\em in vivo} protocol will be hopefully performed in order to validate the numerical
results: this would also be a decisive aid in the transfer of this medical approach from labs to
everyday clinics.

\bibliographystyle{apalike}
\bibliography{paper}

\begin{thebibliography}{}

\bibitem[Aihara and Miyazaki, 1998]{aihara}
Aihara, H. and Miyazaki, J.-i. (1998).
\newblock Gene transfer into muscle by electroporation in vivo.
\newblock {\em Nature Biotechnology}, 16:867--870.

\bibitem[Akerstrom et~al., 2015]{akerstrom}
Akerstrom, T., Vedel, K., Needham, J., and Hojman, P. (2015).
\newblock {Optimizing hyaluronidase dose and plasmid DNA delivery greatly
  improves gene electrotransfer efficiency in rat skeletal muscle}.
\newblock {\em Biochemistry and Biophysics Reports}.

\bibitem[Andr{\'e} and Mir, 2004]{andrereview}
Andr{\'e}, F.~M. and Mir, L.~M. (2004).
\newblock Dna electrotransfer: its principles and an updated review of its
  therapeutic applications.
\newblock {\em Gene therapy}, 11 Suppl 1:S33--42.

\bibitem[Barthelemy and Haftka, 1993]{Barthelemy1993}
Barthelemy, J. F.~M. and Haftka, R.~T. (1993).
\newblock Approximation concepts for optimum structural design - a review.
\newblock {\em Structural optimization}, 5(3):129--144.

\bibitem[Barton and Meckesheimer, 2006]{Barton2006}
Barton, R.~R. and Meckesheimer, M. (2006).
\newblock Chapter 18 metamodel-based simulation optimization.
\newblock In Henderson, S.~G. and Nelson, B.~L., editors, {\em Simulation},
  volume~13 of {\em Handbooks in Operations Research and Management Science},
  pages 535 -- 574. Elsevier.

\bibitem[Baxter and Jain, 1989]{baxterjain}
Baxter, L.~T. and Jain, R.~K. (1989).
\newblock Transport of fluid and macromolecules in tumors. i. role of
  interstitial pressure and convection.
\newblock {\em Microvascular research}.

\bibitem[Buhren et~al., 2016]{buhren}
Buhren, B.~A., Schrumpf, H., Hoff, N.-P., B{\"o}lke, E., Hilton, S., and
  Gerber, P.~A. (2016).
\newblock {Hyaluronidase: from clinical applications to molecular and cellular
  mechanisms.}
\newblock {\em European journal of medical research}, 21(1):5.

\bibitem[Bureau et~al., 2004]{bureau}
Bureau, M.~F., Naimi, S., Ibad, T.~R., and Seguin, J. (2004).
\newblock Intramuscular plasmid dna electrotransfer: biodistribution and
  degradation.
\newblock {\em Biochimica et Biophysica Acta (BBA)}.

\bibitem[Chiarella et~al., 2013a]{Chiarella2013a}
Chiarella, P., De~Santis, S., Fazio, V.~M., and Signori, E. (2013a).
\newblock Hyaluronidase contributes to early inflammatory events induced by
  electrotransfer in mouse skeletal muscle.
\newblock {\em Human Gene Therapy}, 24(4):406--416.

\bibitem[Chiarella et~al., 2013b]{Chiarella2013b}
Chiarella, P., Fazio, V.~M., and Signori, E. (2013b).
\newblock Electroporation in dna vaccination protocols against cancer.
\newblock {\em Current Drug Metabolism}, 14(3):291--299.

\bibitem[Chiarella and Signori, 2014]{Chiarella2014}
Chiarella, P. and Signori, E. (2014).
\newblock Intramuscular dna vaccination protocols mediated by electric fields.
\newblock {\em Electroporation Protocols}, 1121:315--324.

\bibitem[De~Robertis et~al., 2018]{Derobertis}
De~Robertis, M., Pasquet, L., Loiacono, L., Bellard, E., Messina, L., Vaccaro,
  S., Di~Pasquale, R., Fazio, V.~M., Rols, M.-P., Teissi{'e}, J., Golzio, M.,
  and Signori, E. (2018).
\newblock In vivo evaluation of a new recombinant hyaluronidase to improve gene
  electro-transfer protocols for dna-based drug delivery against cancer.
\newblock {\em Cancers}, 10(11):1--20.

\bibitem[Deville, 2017]{Manon}
Deville, M. (2017).
\newblock {\em Mathematical mode of enhanced grug deliery by mean of
  electroporation or enzymatic treatment}.
\newblock PhD thesis, Universit\'e de Bordeaux \& Universit\`a di Roma Tor
  Vergata.

\bibitem[Deville et~al., 2018]{Deville2018}
Deville, M., Natalini, R., and Poignard, C. (2018).
\newblock A continuum mechanics model of enzyme-based tissue degradation in
  cancer therapies.
\newblock {\em Bulletin of Mathematical Biology}, 80(12):3184--3226.

\bibitem[Girish and Kemparaju, 2007]{girish}
Girish, K.~S. and Kemparaju, K. (2007).
\newblock The magic glue hyaluronan and its eraser hyaluronidase: a biological
  overview.
\newblock {\em Life sciences}, 80(21):1921--1943.

\bibitem[Grazia et~al., 2014]{Notarangelo2014}
Grazia, N.~M., Roberto, N., and Emanuela, S. (2014).
\newblock Gene therapy: the role of cytoskeleton in gene transfer studies based
  on biology and mathematics.
\newblock {\em Current Gene Therapy}, 14(2):121--127.

\bibitem[Hecht, 2012]{FreeFem}
Hecht, F. (2012).
\newblock New development in freefem++.
\newblock {\em Journal of Numerical Mathematics}, 20(3-4):251--265.

\bibitem[Hedayat et~al., 1999]{Sloane}
Hedayat, A.~S., Sloane, N. J.~A., and Stufken, J. (1999).
\newblock {\em Orthogonal Arrays: Theory and Applications}.
\newblock Springer-Verlag, New York.

\bibitem[Lang et~al., 2016]{lang}
Lang, G.~E., Vella, D., Waters, S.~L., and Goriely, A. (2016).
\newblock {Mathematical modelling of blood-brain barrier failure and oedema.}
\newblock {\em Mathematical medicine and biology : a journal of the IMA}.

\bibitem[Legu\`ebe et~al., 2017]{Leguebe2017}
Legu\`ebe, M., Notarangelo, M.~G., Twarogowska, M., Natalini, R., and Clair, P.
  (2017).
\newblock Mathematical model for transport of dna plasmids from the external
  medium up to the nucleus by electroporation.
\newblock {\em Mathematical Biosciences}, 285(Supplement C):1 -- 13.

\bibitem[McMahon et~al., 2001]{signori}
McMahon, J.~M., Signori, E., Wells, K.~E., Fazio, V.~M., and Wells, D.~J.
  (2001).
\newblock Optimisation of electrotransfer of plasmid into skeletal muscle by
  pretreatment with hyaluronidase - increased expression with reduced muscle
  damage.
\newblock {\em Gene Therapy}, 8:1264--1270.

\bibitem[Peri, 2009]{peri2009self}
Peri, D. (2009).
\newblock Self-learning metamodels for optimization.
\newblock {\em Ship Technology Research}, 56(3):95--109.

\bibitem[Peri, 2018]{peri2017spline}
Peri, D. (2018).
\newblock Easy-to-implement multidimensional spline interpolation with
  application to ship design optimisation.
\newblock {\em Ship Technology Research}, 65(1):32--46.

\bibitem[Rols et~al., 1998]{rols}
Rols, M., Delteil, C., Golzio, M., Dumond, P., Cros, S., and J, T. (1998).
\newblock In vivo electrically mediated protein and gene transfer in murine
  melanoma.
\newblock {\em Nature Biotechnology}, 16(2):168--171.

\bibitem[Schertzer et~al., 2006]{schertzer}
Schertzer, J.~D., Plant, D.~R., and Lynch, G.~S. (2006).
\newblock Optimizing plasmid-based gene transfer for investigating skeletal
  muscle structure and function.
\newblock {\em Molecular Therapy}, 13(4).

\bibitem[Shu et~al., 2017]{Shu2016}
Shu, L., Jiang, P., Wan, L., Zhou, Q., Shao, X., and Zhang, Y. (2017).
\newblock Metamodel-based design optimization employing a novel sequential
  sampling strategy.
\newblock {\em Engineering Computations}, 34(8):2547--2564.

\bibitem[Soltani and Chen, 2012]{soltani}
Soltani, M. and Chen, P. (2012).
\newblock Effect of tumor shape and size on drug delivery to solid tumors.
\newblock {\em Journal of biological engineering}.

\bibitem[Swartz and Fleury, 2007]{swartz}
Swartz, M.~A. and Fleury, M.~E. (2007).
\newblock {Interstitial flow and its effects in soft tissues.}
\newblock {\em Annual review of biomedical engineering}, 9:229--256.

\bibitem[Toropov, 1995]{Toropov1995}
Toropov, V.~V. (1995).
\newblock {\em Multipoint Approximation Method for Structural Optimization
  Problems with Noisy Function Values}, pages 109--122.
\newblock Springer Berlin Heidelberg, Berlin, Heidelberg.

\bibitem[Ward and Lieber, 2005]{ward}
Ward, S.~R. and Lieber, R.~L. (2005).
\newblock {Density and hydration of fresh and fixed human skeletal muscle.}
\newblock {\em Journal of biomechanics}, 38(11):2317--2320.

\bibitem[Wolff et~al., 1990]{wolff}
Wolff, J.~A., Malone, R.~W., Williams, P., Chong, W., Acsadi, G., Jani, A., and
  Felgner, P.~L. (1990).
\newblock Direct gene transfer into mouse muscle in vivo.
\newblock {\em Science (New York, N.Y.)}, 247(4949 Pt 1):1465--1468.

\bibitem[Yao et~al., 2012]{weiyao}
Yao, W., Li, Y., and Ding, G. (2012).
\newblock {Interstitial fluid flow: the mechanical environment of cells and
  foundation of meridians.}
\newblock {\em Evidence-based complementary and alternative medicine : eCAM},
  2012:853516.

\bibitem[Z{\"o}llner et~al., 2012]{zollner}
Z{\"o}llner, A.~M., Abilez, O.~J., B{\"o}l, M., and Kuhl, E. (2012).
\newblock Stretching skeletal muscle: chronic muscle lengthening through
  sarcomerogenesis.
\newblock {\em PloS one}, 7(10):e45661.

\end{thebibliography}

\end{document}